\documentstyle[12pt,aaspp4]{article}
\def\clock{\count0=\time \divide\count0 by 60
     \count1=\count0 \multiply\count1 by -60 \advance\count1 by \time
     \number\count0:\ifnum\count1<10{0\number\count1}\else\number\count1\fi}

\begin{document}
\title{Heating of the IGM}
\author{Ue-Li Pen}
\affil{Canadian Institute for Theoretical Astrophysics,
60 St George St., Toronto, Ont. M5S 3H8, Canada \\
and Harvard-Smithsonian Center for Astrophysics}

\newcommand{\etal}{{\it et al. }}
\newcommand{\beq}{\begin{equation}}
\newcommand{\eeq}{\end{equation}}

\begin{abstract}

Using the cosmic virial theorem, Press-Schechter analysis and numerical
simulations, we compute the expected X-ray background (XRB) from the diffuse
IGM with the clumping factor expected from gravitational shock
heating.  The predicted fluxes and temperatures are excluded from the
observed XRB.  The predicted clumping can be reduced by
entropy injection.  The required energy is computed from the two-point
correlation function, as well as from Press-Schechter formalisms.  The
minimal energy injection of 1 keV/nucleon excludes radiative or
gravitational heating as a primary energy source.  We argue that the
intergalactic medium (IGM) must have been heated through violent
processes such as massive supernova bursts.  If the heating proceeded
through supernova explosions, it likely proceeded in bursts which may
be observable in high redshift supernova searches.
Within our model we reproduce the observed cluster
luminosity-temperature relation with energy injection of 1 keV/nucleon
if this injection is assumed to be uncorrelated with the local
density.  These parameters predict that the diffuse IGM soft XRB has a
temperature of $\sim$ 1 keV with a flux near 10 keV/cm$^2$ s str keV,
which may be detectable in the near future.

\end{abstract}

{\it Subject Headings:} cosmology: theory and diffuse radiation --
X-rays: general -- supernova: general

\section{Introduction}
The intergalactic medium has been a mysterious subject for many
decades.  While the baryonic content of the universe is still
uncertain, the current Big-Bang nucleosynthesis value of $\Omega_b
h^2=0.02$ (Burles and Tytler 1998) is much larger than the observed
stellar density $\Omega_*h=0.0025$ (Fukogita \etal 1997).  In recent
years, the large gas mass in clusters of galaxies (White \etal 1993)
has uncovered a diffuse reservoir of baryons.  Since the gas fraction
in clusters should be representative of the global mean, this hot
diffuse component of baryons could account for the majority of baryons
in the universe.
Barcons \etal (1991) first applied the soft X-ray background (XRB)
constraints to the IGM.  Since their work, the major portion of the
XRB has been resolved into point sources (Hasinger \etal 1993),
strengthening the bound.  We here present new aspects to the argument
by estimating the temperature with the cosmic virial theorem, and
calculating the clumping factor directly through numerical simulations
and Press-Schechter estimates.  These two ingredients rule out a
passively evolving IGM where the energetics are dominated by gravity.
We discuss these constraints in more detail in section \ref{sec:xrb}.
Using our galaxy as a representative gravitational well, we place
constraints on the non-gravitational heat injection required to
explain the small observed XRB contribution from our halo in section
\ref{sec:galaxy}. In section \ref{sec:test} we discuss predictions and
implications of the model.  It explains the luminosity-temperature
relation in clusters of galaxies, and predicts a very sudden heating
period for each galaxy.
\section{Cosmological X-ray background}
\label{sec:xrb}
The cosmological background flux is
\beq
F=\Omega_b^2\rho_c^2\epsilon_{\rm ff, bol}\int_0^\infty 
\frac{T^{1/2} \langle\delta^2\rangle (1+z)^3}{4\pi d_L^2}d_A^2
\frac{dr}{dz} dz \ \ \frac{\rm erg}{{\rm cm}^2\ {\rm s\ str}}
\label{eqn:bkgnd}
\eeq
where $d_A$ and $d_L$ denote the angular diameter and luminosity
distance respectively, related by $d_L=(1+z)^2 d_A$.  $\rho_c\equiv
3H_0^2/8\pi G$ is the critical closure density of the universe.
$\delta$ is the local overdensity of the gas which is related to  
$\langle\delta^2\rangle=\xi(0)$ in terms of the correlation function at
zero lag.  The volume emission coefficient
$\epsilon_{\rm ff, bol} \sim 1.7 \times 10^{-42} c_Z /0.875 m_p^2$
(\cite{ryb79}) is expressed in terms of the proton mass
$m_p$ at a solar hydrogen-helium mixture with a coefficient $c_Z \sim
Z/Z_\odot(4 {\rm keV}/T)+1$ to correct for metal cooling (Raymond \etal
1976).  The approximation for $c_Z$ is accurate to 20\% in the
interval 0.01 keV 
$<$ T $<$ 10 keV.  We have assumed a mean gaunt factor $g_{\rm ff}\sim
1.2$.   We will use the $Z=Z_\odot/4$ as the lower plausible limit on
the intergalactic medium inferred from clusters of galaxies.   Higher
metallicities strengthen the bounds derived here.

In a flat universe with a powerlaw correlation function
$\xi=(r/r_0)^{-1.8}/(1+z)^2$ the characteristic temperature evolves as
$T^{1/2}\propto (1+z)^{-0.6}$, and is expected to evolve more slowly
in a low density model.  To simplify (\ref{eqn:bkgnd}), we neglect
this weak redshift dependence.  We
express the integrand in terms of $\tau_0\equiv t_0 H_0$, the age
of the universe in units of the Hubble constant.  Recall that
$\tau_0=2/3$ for a flat universe and $\tau_0=1$ for an empty universe.
Since $F=\xi(0)\Omega_b^2\rho_c^2\epsilon_{\rm ff}\tau_0 {c}/{H_0}$,
we find in differential energy units 
\begin{equation}
F= \xi(0)
\left(\frac{\Omega_b h^2}{0.02}\right)^2 \left( \frac{T}{0.13 \rm
\ keV}\right )^{-1/2} \frac{\tau_0}{h} \exp(-\frac{h\nu}{kT})
\left(\frac{\rm keV}{T}+1\right) \frac{0.16
\rm \ keV}{\rm cm^2\ s\ str\ keV} .
\label{eqn:f}
\end{equation}
X-ray emission is proportional to the square of the density.  Its
emission weighted temperature
is well approximated by the pair weighted temperature
given be the cosmic virial theorem (Peebles 1980)
$
\langle T^{1/2}\rangle = \frac{\sigma_8\Omega^{0.5}}{0.53}
\left(\frac{r_g}{100\ \rm kpc}\right)^{0.2} \times 0.128\ \rm keV
$
where $r_g$ is a short scale cutoff of the gas autocorrelation
function. 
From the cluster abundance (Pen 1998b), we use
$\sigma_8=.53 \Omega^{-.55}$, and use
the approximation $\langle T \rangle_X = 0.13$ keV.

The observed extragalactic XRB at 1/4 keV has an upper
limit of 15 keV/cm$^2$ s keV str (Cui \etal 1996).  From shadowing
studies, they found an extragalactic background of 45 keV/cm$^2$ s
keV str, of which at least 30 keV/cm$^2$ s keV str has been resolved
into point sources (Hasinger \etal 1993).  Using $\tau_0/h=1$ and
taking into account a 
factor of 2 suppression of the flux due to the energy dependent Gaunt
factor (Rybicki and Lightman 1979), we  obtain the main constraint 
\beq
\xi(0)<60.
\label{eqn:xi}
\eeq
A less certain prediction is the clumping factor $\xi(0)$.  For a
scale invariant spectrum in a flat universe, this is expected to be
constant in time.  According to Press-Schechter theory, most of the
mass is in collapsed and virialized objects of approximately the
non-linear mass scale.  The average overdensity of collapsed objects
is at least 178, and generally higher if the objects are not
homogeneous or have collapsed at higher redshift.  We expect
at least a factor of two higher clumping, since virialized objects are
certainly not uniform density objects.  
Estimates using stable clustering (Jain 1997) and numerical pure
N-body simulations indicate that at least for the dark matter
component, $\xi(0) \ga 10^4$.  With these values, we far exceed the
observed diffuse soft XRB constraints.
If the distribution of gas
traces light down to scale $r_g$, and then forms a constant density
core, we find $\xi(0) = (r_g/5.5 h^{-1} {\rm Mpc})^{-1.8}$.  Using
(\ref{eqn:xi}) we find $r_g>800 h^{-1}$ kpc, i.e. gas does not cluster
on scales shorter than a Mpc, even though all other forms of matter
do.  

We can compare these estimates with direct numerical simulations (Cen
\etal 1995).  Scaling their Figures 3 and 4 to the cosmological XRB
by multiplying the vertical abscissa by $5\times 10^{38}$
to obtain units of keV/cm$^2$\ s\ str\ keV, we find that they predict
emission just below the observed upper bounds.  Emission is weighted
by the smallest scale clumping, which simulations at finite resolution
always underestimate.  We can see the effect in their Figure 8.  Their
COBE normalized simulation should predict $\rho^2$ weighted
temperatures of $\sim 0.3$ keV, while the simulated temperatures are
signficantly higher.  Increasing the resolution would primarily
enhance the emission of smaller mass and lower temperature objects,
which are smaller and more subject to numerical limitations.  It would
also raise the clumping factor and X-ray emission.  More recent
simulations may move the results in the expected direction (Cen and
Ostriker 1998).

\begin{figure}
\plotone{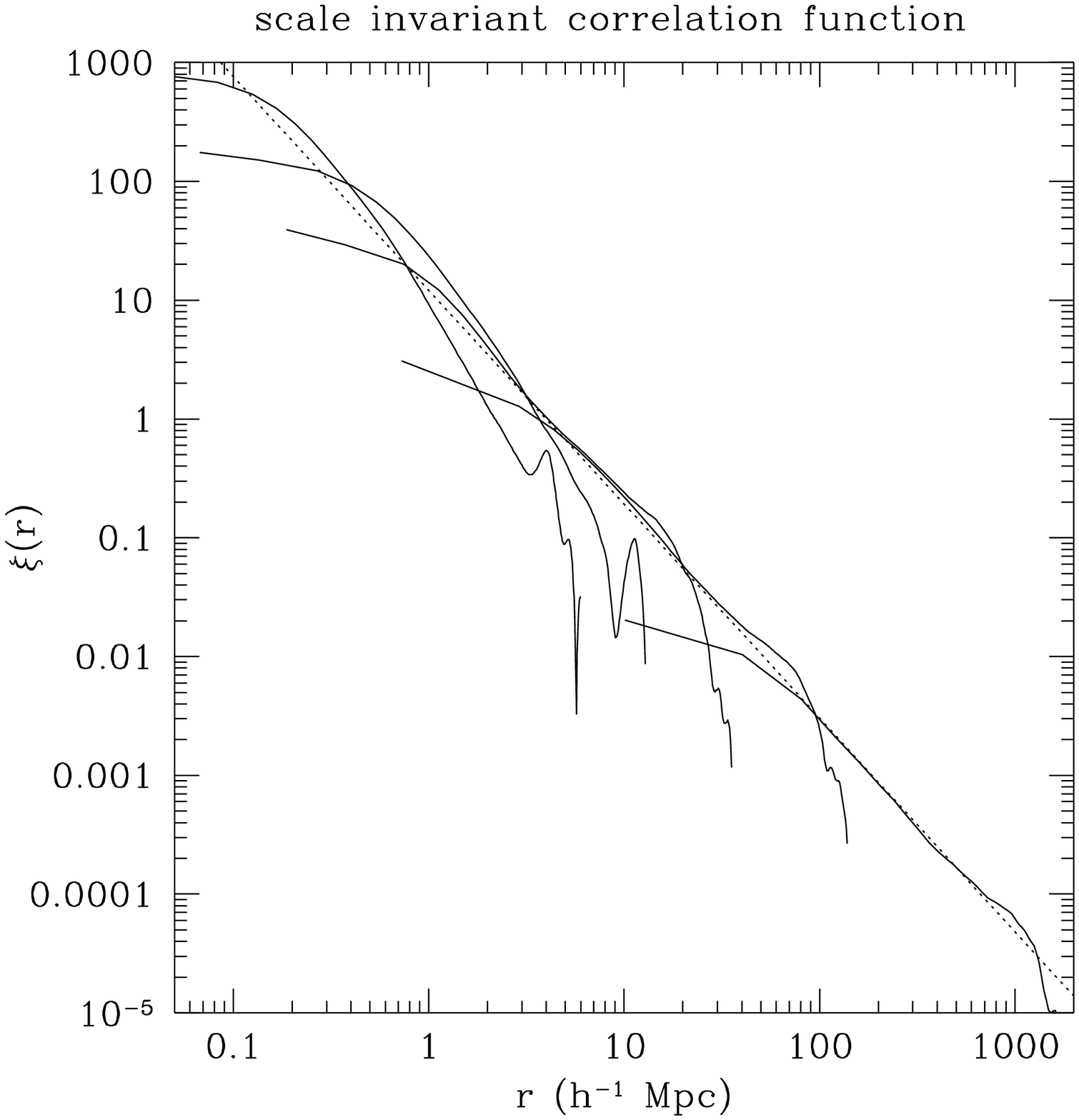}
\caption{
Numerical estimates of the correlation function clumping factor
$\langle C\rangle =\xi(0)$ for gas in adiabatic scale free
simulations.  The solid lines show the correlation function in the
simulation at redshifts $z=180,16,4,1,0$ scaled to the appropriate
non-linear mass scale.  The dashed line is the scale free correlation
for $n=-1.8$ and $r_0=4\ h^{-1}$Mpc.  We see that the simulations are
still limited by resolution, and give a lower bound of $\langle
C\rangle\gtrsim 900$.  This rules out passively evolving IGM models,
and requires significant active heating to evade soft XRB
constraints.  }
\label{fig:corr}
\end{figure}

Using the cosmological Moving Mesh Hydro Code (Pen 1998a), we have
performed direct simulations to measure $\xi(0)$.  We use a scale-free
CDM models with $128^3$ grid cells and an equal number of dark matter
particles.  The correlation function is $\xi\propto (r/r_0)^{-1.8}$
with $r_0=4\ h^{-1}$Mpc in a 16 $h^{-1}$Mpc box.  In order to extend
the dynamic range, the initial conditions were determined by factoring
the correlation function (Pen 1997).  The grid was allowed to compress
a factor of 10, giving a minimal grid spacing of $13 h^{-1}$ kpc, with
an effective resolution which is several times that value.  We used
$\Omega_b=0.1$.  In Figure \ref{fig:corr} we plot the gas correlation
function $\xi(r)$ at different redshifts scaled to their relative
amplitudes using scale invariance.  The simulation moves from the
lower right to the upper left as the non-linear mass scale grows, and
the effective resolution increases.  We see that the clumping factor
is limited by the initial resolution at all times.  The clumping
factor $\xi(0)$ monotonically increases with time, reaching a value of
$\sim 900$ at $z=0$ when the non-linear mass scale reaches 32 grid
cells.  Our limited resolution allows us to only place a lower bound
on $\xi(0)\gtrsim 900$.  One would expect low density models to have
higher clumping factors since objects formed earlier.

We must thus conclude that the gas could not have followed the
gravitational evolution of the dark matter, as one would have expected
in the scenario of purely gravitationally driven collapse.  Two
potential solutions come to mind.  Baryons could have cooled into
compact objects which do not contribute to the XRB.  But this poses
problems for clusters of galaxies, where vast quantities of gas have
been observed (White \etal 1993).  In a hierarchical model of structure
formation, clusters form from the merger of smaller objects.  But if
the smaller objects had their baryons locked up into compact objects,
it would be very difficult to release the gas again to form the
Intra-cluster medium (ICM).

The second solution would be to heat the gas sufficiently, such that
it does not fall into the gravitational potential well of the dark
matter halos.  The required energies can be estimated from the
two-point correlation function.  We calculate the energy required to
convolve the gas with a spherical tophat to satisfy (\ref{eqn:xi}).
For a gas distribution $\rho_g$ in a dark matter potential $\phi$, we
have $E=\int \rho_g \phi$, which for our power law correlation
formally diverges due to long wavelength contributions.  We consider a
specific model in which a smoothed density distribution is given as a
convolution over the original gas field.  The change in energy is
finite, in fact $\Delta E=\int \Delta \rho_g \phi$ and the average
specific energy change is $\Delta {\cal E} = \Omega_0 \rho_c 4\pi G
r_0^2 \int W(r/r_0)\xi(r/r_0)d^3(r/r_0)$.  We let $W$ be a spherical
tophat of radius $r_g$.  Using $\sigma_8$ from above, we obtain
$\Delta {\cal E} = 0.8\times 3 (5-n)^{-1}(3-n)^{-1}(2-n)^{-1}
(r_g/r_0)^{2-n}$ keV, where $n\sim 1.8$.  The constraint for the
smoothed gas field simply translates to $\sigma_{r_g}^2 \lesssim 60$,
which can again be expressed in terms of the correlation function $r_g
\gtrsim 823 \sigma_8^{10/9} h^{-1}$ kpc.  In this model, approximately
2 keV of heating would be required to satisfy the constraints from the
XRB.  This assumes that smoothing was applied to all points in space,
which probably overestimates the energy requirement.  We obtain a
lower energy bound by just truncating the correlation function, so
that $\xi(r)=(r_g/r_0)^{-1.8}$ for $r<r_g$.  This requires $0.25$ keV,
but must be taken as a lower limit since it violates locality.
We will now proceed to build a more realistic model which allows us to
quantify the heat injection required to evade the bound from equation
(\ref{eqn:xi}).  
\section{Halo Model}
\label{sec:halo}
We will consider a model for
the distribution of gas within each halo.  We assume the
gravitational potential to be dominated by dark matter with a singular
isothermal sphere mass distribution.  The mass-rotation relation for
the dark matter is taken from the isothermal Press-Schechter model
(Eke, Cole and Frenk 1997)
$M=\frac{v_c^3}{G^{3/2}}\left(\frac{3}{4\pi \Delta \rho_c}\right)^{1/2}.
$ 
$v_c$ is the circular rotation speed of the object.
The virial radius $r_\Delta$
of an object is defined as the fiducial point where
the mean interior density is $\Delta\sim 178\Omega^{-1/2}$ times the mean
cosmological density.  It corresponds to half the turnaround radius of
a tophat model.  For an isothermal sphere
$
r_\Delta=v_c\sqrt{\frac{3}{4\pi G\rho_c\Delta}}.
$ 
The gas is assumed to have two phases: an outer region $r>r_1$ where gas
traces mass isothermally with a gas fraction $f_g$, 
and an inner region where the gas has been heated to
constant entropy.  Such a distribution is the generic outcome of
central non-gravitational heating where entropy is injected in the
center, which is then convectively transported outward.
Observationally (Danos and Pen 1998, Cooray 1998),
the best fit for the gas fraction is $f_g=0.06 h^{-3/2}$.
For the outer region, the gas density profile is
given by an isothermal sphere $\rho_g=f_gv_c^2/4\pi r^2G$.  In the
inner region $r<r_1$, the gas density is then
$
\rho_g=\frac{v_c^2f_g}{4\pi r_1^2 G}\left[1+\frac{12}{25}\log(\frac{r_1}{r})
\right]^{3/2}.
$
We can define an equivalent core radius $r_c=0.487r_1$.  For this
choice, the free-free luminosity of the two phase object is identical
to an isothermal $\beta=2/3$ model with core radius $r_c$ (Jones and
Forman 1984).
$
L_X=\epsilon_{\rm ff}{f_g^2 v_c^4 T^{1/2}}/({16 r_c G^2}).
$
Due to the enhanced central entropy,
the emission weighted temperature of our objects will be 10\% higher
than the isothermal temperature 
$
\langle T\rangle_X=1.1\times \mu m_p v_c^2/2k.
$
We can estimate the energy required to increase $r_1$.  We
calculate the difference in binding energy between a singular
isothermal gas distribution in the dark matter well, to one with
finite $r_1$.  While the energy injection in our model is concentrated
to the center, we define a mean energy $\delta T$ to be the total
injected energy divided by the total virial gas mass.  We obtain 
$
\frac{r_1}{r_\Delta}=1.582\left(1-\sqrt{1-1.029\frac{\delta T}{T}}\right)
$
In the limit of $\delta
T/T\ll 1$, we find 
$
\frac{r_1}{r_\Delta} = 0.814 \frac{\delta T}{T}.
$
The clumping factor $C$ of an
individual object is 
$
C = \frac{\Delta}{3}\left(1.553\frac{r_\Delta}{r_1}-1\right)
$
and in the limit that $T_1 \ll T$, we have the linear relation
$C=0.52 T \Delta/\delta T$.  We find the average clumping factor
$
\langle C\rangle=\xi(0)=\int_{\delta T}^\infty\frac{df}{dT} C(T) dT
$
where the lower cutoff assumes that objects with energy
injection larger than their virial temperature have ejected all their
gas.  The Press-Schechter distribution function is 
$
f(>M) = \sqrt{\frac{2}{\pi}} \int_\frac{\delta_c}{\sigma}^\infty
e^{-u^2/2} du.
$
We use the standard value $\delta_c=1.686$.
Using a power law mass fluctuation spectrum
$\sigma(M)=(M/M_8)^{-\alpha}$ (Pen 1998b), we find
\beq
\langle C
\rangle = \frac{3}{2} \sqrt{\frac{2}{\pi}} \frac{T_8}{\delta T}0.518
\alpha \delta_c 
\Delta \left( \frac{2 \sigma_8}{\delta_c}\right)^{\frac{2}{3\alpha}-1}
\int_{w_1}^\infty w^{\frac{1}{3\alpha}-\frac{1}{2}} e^{-w} dw
\label{eqn:cr}
\eeq
where $T_8\sim 5 \Omega_0^{2/3}$ (Pen 1998b) and  $w_1=\delta_c^2
(\delta T/T_8)^{3\alpha}/2\sigma_8^2$.   
For most CDM-like spectra, $\alpha \sim 1/3$, which simplifies the
(\ref{eqn:cr}) to
$
\langle C \rangle=\frac{100}{\Omega^{1/3}}\times
\left[
2\sqrt{w_1}e^{-w_1}+\sqrt{\pi}{\rm erfc}(\sqrt{w_1})\right]
\frac{\rm keV}{\delta T}
$
where now $w_1=5\Omega \delta T/T$.  The leading order expansion of
the bracket in is $\sqrt{\pi}- (4w_1^{3/2}/3)$.
We can compare the predicted
clumping to the constraint (\ref{eqn:f}). 
By assumption, all gas has a temperature of $\gtrsim \delta T$.
For an $\Omega=1$ scenario, we
can satisfy the diffuse XRB constraint at 1 keV  for $\delta T
\gtrsim 1$ keV. We used an extragalactic diffuse flux limit of 8
keV/cm$^2$ s str keV (Hasinger \etal 1993) at 1 keV.
\section{Galactic X-ray background}
\label{sec:galaxy}
While the cosmological constraint depends on parameters including
geometry of the universe and the evolution of the clumping, our own
Galaxy provides a completely independent constraint on the thermal
state of the intergalactic medium.  Simulations (Navarro
\etal 1995) and
observations indicate that the matter distribution is well
approximated by an isothermal sphere extending from ca 1/10th of the
virial radius out to the virial radius.  In that range, gas and dark
matter do appear 
to trace each other to within a factor of two, and are within a factor
of two of the singular isothermal sphere. 
Approximating our solar system to be at the
center of the gravitational galaxy halo, we find
\beq
F=\epsilon T^{1/2}r_1 \left(\frac{v_c^2 f_g}{4\pi
G r_1^2}\right)^2 \left[\frac{1}{3}+ \int_0^1 (1-\frac{12}{25}\log
u)^{7/2} du  \right] \ \ \frac{\rm erg}{{\rm cm}^2\ {\rm s\ str}} .
\label{eqn:gal}
\eeq
To support the outer gas in pressure equilibrium, the central gas will
have a temperature higher than the virial halo temperature.
The emission weighted temperature seen at the center differs from
the virial temperature, becomes instead $T_X=2.25\times \mu m_p
v_c^2/2k$, more than twice the isothermal value.
For a rotation speed of 220 km/s, we expect $T_X=0.34
$keV, independent of the core radius $r_1$ as long it is greater than
the heliocentric distance: $r_1 \gg R_0$.  

The 1/4 keV galactic background is known to be dominated by the local
hot ISM bubble, which is difficult to disentangle from a halo
distribution.  Instead, we will use the 1 keV background constraint,
for which (\ref{eqn:gal}) yields $F=0.003/(h^3 r_1^3)$ keV/cm$^2$ s
str keV.  $r_1$ is measured in Mpc.  Current reports lie near 20
keV/cm$^2$ s str keV (Miyaji \etal 1998), of which Hasinger \etal
(1993) 
constrain the homogeneous component to less than 20\%.  This value is
lower than the previously used value since the cosmological XRB
could be clumped.  
The lower limit for the heated core radius is
$
r_1>150 h^{-1} \rm kpc.
$
The virial radius of our galaxy is $r_\Delta = 127 h^{-1}$ kpc,
which is smaller than the required core radius.  This corresponds to a
scenario where all the IGM gas must have been completely expelled from
our galaxy.  Unbinding the gas from the galactic potential well
requires a minimum heat injection of 150 eV per nucleon.  We now have
two independent lines of evidence that very significant
non-gravitational heating of the IGM must have occurred.  Supernovae
might be one potential energy source which we will discuss further in
section \ref{sec:test}.

In the calculation of our energy injection estimates, we have taken
the binding energy difference between a model where the gas traces the
mass to one where it satisfies the XRB constraint.  It
would seem plausible that this is a lower bound to the actual energy
injection required.  In addition to the potential energy change, a
similar amount of energy may be expended to raise the entropy of the
gas.  Placing the heating process too early causes some of the
injected energy to be lost to cosmic expansion, thus requiring even
larger energy inputs.  We should also note that in a cosmological
context, energy is not strictly conserved, and it is in principle
possible to extract energy from the dark matter through its time
dependent potential.  Simulations seem to suggest that gas traces the
total mass up to overdensities of at least $10^3$ (Navarro \etal
1995), suggesting that in fact this energy exchange process does not
play a dominant role.  We assumed that objects
formed recently.  An earlier formation time would increase the
clumping factor, and raise the predicted X-ray emission.  
\section{Potential Tests}
\label{sec:test}
We now consider several predictions of the violent heating model.  We
have assumed an equal specific energy injection in all objects
independent of mass.  This model predicts the core radius $r_c \propto
T^{-1}$, which can be compared to the cluster luminosity-temperature
relation.  From the equations in section \ref{sec:halo}, 
$
L_{44} = 2.1 h^{-2}\Omega^{-1/4}\left(\frac{T}{\rm 5\ keV}
\right)^3\left(\frac{\rm 
keV}{\delta  T}\right) \times 10^{44} \rm erg/s
$
which can be compared to the Henry and Arnaud (1991) data.  Their
relation is consistent with the slope $L\propto T^3$,
and agrees for values of $ \delta T \sim$ 1 keV.
Metzler and Evrard (1994) had studied numerical simulations with similar
parameters and arrived at consistent conclusions.  In this simplest
model where the heating is uncorrelated with the mass of the
object, we find that the required heating to explain the cluster core
radii can simultaneously explain the soft XRB.

If the current spatial stellar distribution is a tracer of the heat
source distribution, we obtain an upper limit on the heating time
scale to be the cooling time of the gas.  The observed metalicities in
clusters is about 1/3 solar, suggesting that approximately 1\% of the
gas had undergone nuclear burning, releasing a few MeV of kinetic
energy for each burnt nucleon.  We have a mean energy injection about
10 keV for the gas, of which 10\% must remain after radiative and
adiabatic losses.  We require that the heating time not be more than
10 times longer than the cooling time in order for the injected heat
to be retained.
The local luminosity density in the disk is $\sim 0.07 L_\odot$/pc$^3$
(Binney and Tremaine 1987).  
Our galaxy has a virial mass of $2.6\times 10^{12}M_\odot$, and using
the cosmological gas fraction $f_g$ from above, we expect a total gas
mass of $1.6\times 10^{11} h^{-3/2}M_\odot$.  At a total luminosity of
$1.4\times 10^{10}L_\odot$, we have a mean gas to light ratio of $11
h^{-3/2}M_\odot/L_\odot$.  We expect a cooling
time of 
$
t_{\rm cool} \sim 2.8\times 10^5\left(\frac{ M_\odot {\rm
pc}^{-3}}{\rho}\right) {\rm years}
$\ assuming a temperature $T=1$ keV.  
Allowing 90\% of the injected
energy to dissipate radiatively, we need to inject the energy in
$\delta t=10 t_{\rm cool}$.  
We infer that
the heating must have occurred over $\sim 10^{7}$ years.

Each supernove injects about ten $M_\odot$ of metals, so we expect
$\sim 10^8$ supernovae in the cooling time.  We draw two conclusions:
1. there must have been a short burst of very active star formation and
supernova explosions, during which about 10 supernovae detonated per
year.  2. the galaxy would have been at least ten times brighter during
this period since only stars of M/L$<$1 will burn out.  The epoch of
this explosive heating is not constrained, and different galaxies may
burst at different times.  We do predict that a small fraction ($\sim
10^{-3}$) of all galaxies at high redshift ($z>1$) undergo enormous
supernova bursts.  Dust obscuration can weaken these luminosities.
\section{Conclusion}
We have shown that the current constraints from the XRB
significantly constrain the current thermal state of the intergalactic
medium.  The absence of significant emission from our galactic halo
places a lower bound of 0.15 keV energy injection.  A stronger
constraint arises from the absence of diffuse intergalactic emission,
for which the emission in enhanced by the clumping factor of the gas.
We found that a mean energy injection of $\delta T\gtrsim 1$ keV is
required.  Photoheating is not a possible solution.  Such a process
would raise the temperature of the gas to $\sim 0.002$ keV.  In the
best possible case, this heating occured at $z \sim 0$, and the
objects instantaneously collapse today.  The XRB constraints require
that the overdensity be less than 100, for which adiabatic compression
can raise the energy to at most 0.04 keV, far short of the required
heat input.
Our halo model also accounts for the observed cluster
luminosity-temperature relation if $\delta T \sim 1$ keV.  The
cosmological diffuse XRB should be detectable soon at
temperatures near 1 keV.  This is consistent with past simulations
(Metzler and Evrard 1994, Suginohara and Ostriker 1998).  If the
heating process traced the present day light distribution, we obtain
upper limits on the heating time scale, which predicts galaxies
undergoing short violent supernova bursts at $z\gtrsim 1$.

{\it Acknowledgements} I would like to thank Simon White, Uros Seljak
and Avi Loeb for helpful discussions.  This work was supported in part
by NASA grant NAG5-7039.  Computing time was provided by the National
Center for Supercomputing applications.  Part of this work was done at
the Institute for Astronomy and Astrophysics of the Academia Sinica in
Taipei, Taiwan, whom I thank for their hospitality.

\end{document}